\documentclass[preprint,prc,aps,groupedaddress,amsfonts]{revtex4}
\usepackage{amssymb}

\usepackage{dsfont}
\usepackage{amsmath}

\renewcommand{\arraystretch}{0.8}
\font\tenfrak=eufm10  
\font\sevenfrak=eufm7 \font\fivefrak=eufm5
\newfam\frakfam
\textfont\frakfam=\tenfrak \scriptfont\frakfam=\sevenfrak
\scriptscriptfont\frakfam=\fivefrak

\DeclareMathOperator{\Tr}{Tr}

\begin{document}

\title{Canonical Quantization of SU(3) Skyrme Model in a General
Representation}

\author{D. Jur\v ciukonis}
\email[]{darius@itpa.lt}
\author{E.Norvai\v sas}
\email[]{norvaisas@itpa.lt}
\affiliation{
Vilnius University Institute of Theoretical Physics and Astronomy,\\
Go\v stauto 12, Vilnius 01108, Lithuania}

\author{D.O. Riska}
\email[]{riska@pcu.helsinki.fi}
\affiliation{Helsinki Institute of Physics and
Department of Physical Sciences\\ 00014 University of Helsinki,
 Finland }

%\date{today}

\begin{abstract}
A complete canonical quantization of the SU(3) Skyrme model
performed in the collective coordinate formalism in general
irreducible representations. In the case of SU(3) the model
differs qualitatively in different representations. The
Wess-Zumino-Witten term vanishes in all self-adjoint
representations in the collective coordinate method for separation
of space and time variables. The canonical quantization generates
representation dependent quantum mass corrections, which can
stabilize the soliton solution. The standard symmetry breaking
mass term, which in general leads to representation mixing,
degenerates to the SU(2) form in all self-adjoint representations.

\end{abstract}

\maketitle

\section{Introduction}

The Skyrme model is a nonlinear field theory, with
localized finite energy soliton solutions, which
may be quantized as fermions \cite{Skyrme61,Skyrme62}.
The semi-classically quantized Skyrme model model has
proven useful for
baryon phenomenology as a realization of the large color
limit of QCD \cite{Adkins}.
The original model was defined for a
unitary field $U(\mathbf{x},t)$
that belongs to
fundamental representation
of SU(2). The boundary condition $U\rightarrow \mathds{1}$ as
$\left| \mathbf{x}\right| \rightarrow \infty $ implies that
the unitary
field represents a mapping from
$S^{3}\rightarrow S^{3}$, the integer valued
winding number of which classifies the solitonic sectors
of the model and may interpreted as the baryon
number.
The model has subsequently been directly
generalized the SU(3) and SU(N) \cite{Walliser}, in which
case the field $U(\mathbf{x},t)$
is described by group valued functions with
semiclassical quantization.

Both the SU(2) and SU(3) Skyrme versions of the model
have been
quantized canonically in refs. \cite{Fujii87, Fujii88}
in the collective coordinate formalism. The
canonical quantization procedure leads to quantum corrections to
the skyrmion mass, which restore the stability of the
soliton solutions that is lost in the semiclassical
quantization. This method has subsequently been generalized
to unitary fields $U(\mathbf{x,}t)$ that belong to general
representations of the SU(2) \cite {Norvaisas, Acus97,
Acus98},
along with a demonstration that the quantum corrections,
which stabilize the soliton solutions, are
representation  dependent.

The aim of the present paper is to extend the
canonically quantized Skyrme model to
general irreducible representations (irrep) of SU(3).
The motivation is the absence of any
{\it a priori} reason to restrict collective chiral
models to the fundamental representation of the group.
The focus here is on the mathematical aspects of the model,
and on the derivation of both the Hamiltonian density and
the Hamiltonian, in order to elucidate their representation
dependence. The possible
phenomenological applications both
in hyperon and hypernuclear phenomenology
as well as in the Skyrme model description of
the quantum
Hall effect \cite{Sondhi} and Bose-Einstein condensates
\cite{Khawaja}, are not elaborated.

In contrast to the case of SU(2), the solutions to
the SU(3) Skyrme model depend
in an essential way on the dimension. Remarkably
the Wess-Zumino-Witten (WZW) term vanishes in all
self-adjoint irreps of SU(3), as it is proportional
to the cubic Casimir operator $C_3^{SU(3)}$
in the collective
coordinate method for
separation of the dependence on space and
time variables. In the self adjoint irreps
the symmetry breaking mass term in the model reduces to the
SU(2) form.

After some preliminary definitions in Section 2 below,
the main part of this paper is organized as follows. In
Section 3, the classical treatment of the Skyrme model in a
general irrep of SU(3) is reviewed. In Section 4, the quantum
Skyrme model is constructed \textit{ab initio} in the collective
coordinates framework. In Section 5, the WZW term
is taken into account and the left and right transformation
generators are derived from the Lagrangian. The Lagrangian and
Hamiltonian density operators are given explicitly in terms of
generators. In Section 6, the symmetry breaking term are
considered in the collective coordinates framework.
Section 7 contains a summarizing discussion. A number of
relevant mathematical
details are given in the Appendix.

\section{Definitions for the unitary SU(3) soliton field}

The unitary field $U(\mathbf{x},t)$ is defined for for general
irreps $(\lambda ,\mu )$ of SU(3) in addition to the fundamental
representation $(1,0)$. The related Young tableaux are denoted
$[\lambda _{1},\lambda _{2},\lambda _{3}]$, where $\lambda
=\lambda _{1}-\lambda _{2}$ , $\mu =\lambda _{2}-\lambda _{3}$. A
group element is specified by the eight real parameters $\alpha
^{i}(\mathbf{x},t)$. The unitary field is expressed in the
form of
Wigner $D$ matrices for SU(3) in $(\lambda ,\mu )$ irrep as:
\begin{equation}
U(\mathbf{x},t)=D^{(\lambda ,\mu )}(\alpha ^{i}(\mathbf{x},t)).  \label{A1}
\end{equation}
The one-form of the unitary field belongs to the Lie algebra
of SU(3).
The one-forms may be determined by the functions
$C_{i}^{(Z,I,M)}(\alpha )$ and $C_{i}^{\prime (Z,I,M)}(\alpha )$,
the explicit expressions for which depend on the specific group
parameterization:
\begin{align}
\partial _{i} U U^\dagger =&\left( \frac{\partial }{\partial
\alpha ^{i}}U\right) U^\dagger =C_{i}^{(Z,I,M)}(\alpha
)\left\langle \hspace{0.2cm}\left| \hspace{0.1cm}J_{(Z,I,M)}^{(1,1)}\hspace{%
0.05cm}\right| \hspace{0.2cm}\right\rangle ,  \notag \\
U^\dagger\partial _{i} U =& \, U^\dagger\frac{%
\partial }{\partial \alpha ^{i}}U=C_{i}^{\prime (Z,I,M)}(\alpha
)\left\langle \hspace{0.2cm}\left| \hspace{0.1cm}J_{(Z,I,M)}^{(1,1)}\hspace{%
0.05cm}\right| \hspace{0.2cm}\right\rangle .  \label{A2}
\end{align}
The parameters spin $I$, and its projections $M$ and $Z$,
which is related to
hypercharge as $Y=-2Z$, specify the basis states of irrep (1,1).

The parameterization for the SU(3) model, and the expressions of
the differential Casimir operator
in terms of the Euler angles, has been proposed by Yabu and
Ando \cite{Yabu}. The SU(3) generators are
defined as components of irreducible tensors $(1,1)$ and may be
expanded in terms of the Gell-Man generators $\Lambda _{k}$:
\begin{alignat}{2}
J_{(0,0,0)}^{(1,1)}=& -\frac{1}{2}\Lambda _{8}\,, & \qquad
J_{(0,1,0)}^{(1,1)}=&
\frac{1}{2}\Lambda _{3}\,, \notag \\
J_{(0,1,1)}^{(1,1)}=& -\frac{1}{2\sqrt{2}}\left( \Lambda
_{1}+i\Lambda _{2}\right) \,, & \qquad J_{(0,1,-1)}^{(1,1)}=&
\frac{1}{2\sqrt{2}}\left(
\Lambda _{1}-i\Lambda _{2}\right) \,, \notag \\
J_{(-\frac{1}{2},\frac{1}{2},\frac{1}{2})}^{(1,1)}=&
\frac{1}{2\sqrt{2}}\left(
\Lambda _{4}+i\Lambda _{5}\right) \,, & \qquad J_{(-\frac{1}{2},\frac{1}{2},-%
\frac{1}{2})}^{(1,1)}=& \frac{1}{2\sqrt{2}}\left( \Lambda
_{6}+i\Lambda
_{7}\right) \,, \notag \\
J_{(\frac{1}{2},\frac{1}{2},\frac{1}{2})}^{(1,1)}=&
-\frac{1}{2\sqrt{2}}\left(
\Lambda _{6}-i\Lambda _{7}\right) \,, & \qquad J_{(\frac{1}{2},\frac{1}{2},-%
\frac{1}{2})}^{(1,1)}=& \frac{1}{2\sqrt{2}}\left( \Lambda
_{4}-i\Lambda _{5}\right) .  \label{A3}
\end{alignat}
In the case of the fundamental representation the
$\Lambda _{k}$
matrices generators reduce to the standard Gell-Mann matrices
$\lambda_i$. Although the generators (\ref{A3}) are non-hermitian: $%
(J_{(Z,I,M)}^{(1,1)}) ^\dagger=(-1)^{Z+M}J_{(-Z,I,-M)}^{(1,1)}$,
the commutation
relations nevertheless have the simple form:
\begin{equation}
\left[ J_{(Z^{\prime },I^{\prime },M^{\prime
})}^{(1,1)},J_{(Z^{\prime \prime },I^{\prime \prime },M^{\prime
\prime })}^{(1,1)}\right] =\sum\limits_{(Z,I,M)}-\sqrt{3}\left[
\begin{array}{ccc}
\scriptstyle(1,1) & \scriptstyle(1,1) & \scriptstyle(1,1)_{a} \\
\scriptstyle(Z^{\prime },I^{\prime },M^{\prime }) &
\scriptstyle(Z^{\prime \prime },I^{\prime \prime },M^{\prime
\prime }) & \scriptstyle(Z,I,M)
\end{array}
\right] J_{(Z,I,M)}^{(1,1)}\, . \label{comm}
\end{equation}
Here the coefficient on the r.h.s. of eq. (\ref{comm}) is a
Clebsch-Gordan coefficient of SU(3), the explicit expressions for
which are given in \cite {Kuriyan}.
The index $a$ in the Clebsch-Gordan
coefficient denotes that only antisymmetric irrep couplings are
included.

For the specification of the basis
states in a general irrep $(\lambda ,\mu )$
the
parameters $(z,j,m)$, where the hypercharge is $y=\frac{2%
}{3}(\mu -\lambda )-2z$, are employed. The basis
state parameters satisfy the
inequalities:
\begin{align}
j-m \geq &0\,,\qquad j-z\geq 0\, ,  \label{A5} \notag \\
j+m \geq &0\,,\qquad j+z\geq 0\, ,  \notag \\
\lambda +z-j \geq &0\,,\qquad \mu -z-j\geq 0\, ,
\end{align}
where the left hand sides are integers.
The
generators (\ref{A3}) act on the basis states as follows:
\begin{align}
J_{(0,0,0)}^{(1,1)} \genfrac{|}{\rangle}{0pt}{}{(\lambda ,\mu )}{
z,j,m} =\hspace{0.03cm}-&\frac{\sqrt{3}}2 y
\genfrac{|}{\rangle}{0pt}{}{(\lambda ,\mu )}{ z,j,m}\,,
\hspace{0.5cm} J_{(0,1,-1)}^{(1,1)}
\genfrac{|}{\rangle}{0pt}{}{(\lambda ,\mu )}{ z,j,m}
=\sqrt{\scriptstyle \frac12(j+m)(j-m+1)}
\genfrac{|}{\rangle}{0pt}{}{(\lambda ,\mu )}{ z,j,m-1}\,,
 \notag \\
J_{(0,1,0)}^{(1,1)} \genfrac{|}{\rangle}{0pt}{}{(\lambda ,\mu )}{
z,j,m} =m& \genfrac{|}{\rangle}{0pt}{}{(\lambda ,\mu )}{ z,j,m}\,,
\hspace{1.42cm} J_{(0,1,1)}^{(1,1)}
\genfrac{|}{\rangle}{0pt}{}{(\lambda ,\mu )}{ z,j,m} =
-\sqrt{\scriptstyle \frac 12(j-m)(j+m+1)}
\genfrac{|}{\rangle}{0pt}{}{(\lambda ,\mu )}{ z,j,m+1}\,,
 \notag \\
 \notag \\
\begin{split}
J_{(-\frac 12,\frac 12,\frac 12)}^{(1,1)}
\genfrac{|}{\rangle}{0pt}{}{(\lambda ,\mu )}{ z,j,m} &
={\textstyle\sqrt{\frac{(\lambda +z-j)(\mu
-z+j+2)(j-z+1)(j+m+1)}{2(2j+1)(2j+2)}}}
\genfrac{|}{\rangle}{0pt}{}{(\lambda
,\mu )}{ z-\frac 12,j+\frac12,m+\frac 12} \notag \\
&-{\textstyle\sqrt{\frac{(\lambda +z+j+1)(\mu
-z-j+1)(j+z)(j-m)}{4j(2j+1)}}}
\genfrac{|}{\rangle}{0pt}{}{(\lambda ,\mu )}{ z-\frac 12,j-\frac
12,m+\frac 12}\,,
\end{split}
 \notag \\
\begin{split}
J_{(\frac 12,\frac 12,-\frac 12)}^{(1,1)}
\genfrac{|}{\rangle}{0pt}{}{(\lambda ,\mu )}{ z,j,m}&
={\textstyle-\sqrt{\frac{(\lambda +z+j+2)(\mu
-z-j)(z+j+1)(j-m+1)}{2(2j+1)(2j+2)}}}
\genfrac{|}{\rangle}{0pt}{}{(\lambda
,\mu )}{ z+\frac 12,j+\frac12,m-\frac 12}  \notag \\
& +{\textstyle\sqrt{\frac{(\lambda +z-j+1)(\mu
-z+j+1)(j-z)(j+m)}{4j(2j+1)}}}
\genfrac{|}{\rangle}{0pt}{}{(\lambda ,\mu )}{ z+\frac 12,j-\frac
12,m-\frac 12}\,,
\end{split}
 \notag \\
\begin{split}
J_{(-\frac 12,\frac 12,-\frac 12)}^{(1,1)}
\genfrac{|}{\rangle}{0pt}{}{(\lambda ,\mu )}{ z,j,m}&
={\textstyle\sqrt{\frac{(\lambda +z-j)(\mu
-z+j+2)(j-z+1)(j-m+1)}{2(2j+1)(2j+2)}}}
\genfrac{|}{\rangle}{0pt}{}{(\lambda
,\mu )}{ z-\frac 12,j+\frac12,m-\frac 12} \notag \\
& +{\textstyle\sqrt{\frac{(\lambda +z+j+1)(\mu
-z-j+1)(j+z)(j+m)}{4j(2j+1)}}}
\genfrac{|}{\rangle}{0pt}{}{(\lambda ,\mu )}{ z-\frac 12,j-\frac
12,m-\frac 12}\,,
\end{split}
 \notag \\
\begin{split}
J_{(\frac 12,\frac 12,\frac 12)}^{(1,1)}
\genfrac{|}{\rangle}{0pt}{}{(\lambda ,\mu )}{ z,j,m}
&=-{\textstyle\sqrt{\frac{(\lambda +z+j+2)(\mu
-z-j)(j+z+1)(j+m+1)}{2(2j+1)(2j+2)}}}
\genfrac{|}{\rangle}{0pt}{}{(\lambda
,\mu )}{ z+\frac 12,j+\frac12,m+\frac 12}  \\
& -{\textstyle\sqrt{\frac{(\lambda +z-j+1)(\mu
-z+j+1)(j-z)(j-m)}{4j(2j+1)}}}
\genfrac{|}{\rangle}{0pt}{}{(\lambda ,\mu )}{ z+\frac 12,j-\frac
12,m+\frac 12}.
\end{split}
\end{align}
The basis states are chosen such that the generators
$J_{(0,1,0)}^{(1,1)}$ and $%
J_{(0,0,0)}^{(1,1)}$, as well as
the Casimir operator of the SU(2) subgroup %
$\hat{C}^{SU(2)}=\sum
(-1)^{M}J_{(0,1,M)}^{(1,1)}J_{(0,1,-M)}^{(1,1)}$, are diagonal and
thus provide a labelling of the basis states:
\begin{equation}
\hat{C}^{SU(2)}\genfrac{|}{\rangle}{0pt}{}{(\lambda ,\mu )}{
z,j,m} = j(j+1)\genfrac{|}{\rangle}{0pt}{}{(\lambda ,\mu )}{
z,j,m} .  \label{A4}
\end{equation}

\section{The Classical SU(3) Skyrme Model}

The action of the Skyrme model in $SU(3)$ is taken to have the form:
\begin{equation}
S=\int \mathrm{d}^{4}x(\mathcal{L}_{Sk}+\mathcal{L}_{SB})+S_{WZ}\, ,
\label{G1}
\end{equation}
where the chirally symmetric Lagrangian density is
\cite{Skyrme61}
\begin{equation}
\mathcal{L}_{Sk}=-\frac{f_{\pi }^{2}}{4}\Tr \{\mathbf{R}_{\mu }\mathbf{R}%
^{\mu }\}+\frac{1}{32\mathrm{e}^{2}}\Tr\{[\mathbf{R}_{\mu },\mathbf{R}_{\nu
}][\mathbf{R}^{\mu },\mathbf{R}^{\nu }]\}\ .  \label{G2}
\end{equation}
Here the right chiral current is defined as
\begin{equation}
\mathbf{R}_{\mu }=\left( \partial _{\mu }U\right)
U^\dagger=\partial _{\mu
}\alpha ^{i}C_{i}^{(A)}(\alpha )\left\langle \hspace{0.2cm}\left| \hspace{%
0.1cm}J_{(A)}^{(1,1)}\hspace{0.05cm}\right| \hspace{0.2cm}\right\rangle .
\label{G3}
\end{equation}
The Greek characters indicate differentiation with respect to
spacetime variables $\partial _{\mu }=\partial/\partial
x^{\mu }$ in the metric $\mathrm{diag}(\eta
_{\alpha \beta })=(1,-1,-1,-1)$. The only parameters of the
model are $f_{\pi }$ and \textrm{e}. The symmetry breaking term
$\mathcal{L}_{SB}$ and Wess-Zumino-Witten action $S_{WZ}$ are
specified below.

Upon substitution of (\ref{G3}) into (\ref{G2}) the
classical Lagrangian density may be expressed in terms of the
group parameters $\alpha ^{i}$ as:
\begin{align}
\mathcal{L}_{Sk} =& \, \frac{3}{32 N}\dim (\lambda ,\mu
)C_{2}^{SU(3)}(\lambda ,\mu )\Bigl\{-f_{\pi }^{2} (-1)^{A}\partial
_{\mu }\alpha ^{i}C_{i}^{(A)}(\alpha )\partial ^{\mu }\alpha
^{i^{\prime
}}C_{i^{\prime }}^{(-A)}(\alpha )  \notag \\
&+\frac{3}{8\mathrm{e}^{2}}\cdot (-1)^{A}\partial _{\mu }\alpha
^{i}C_{i}^{(A^{1})}(\alpha )\partial _{\nu }\alpha ^{i^{\prime
}}C_{i^{\prime }}^{(A^{2})}(\alpha )  \notag \\
&\times \partial ^{\mu }\alpha ^{i^{\prime \prime }}C_{i^{\prime
\prime }}^{(A^{3})}(\alpha )\partial ^{\nu }\alpha ^{i^{\prime
\prime \prime
}}C_{i^{\prime \prime \prime }}^{(A^{4})}(\alpha )\renewcommand{%
\arraystretch}{0.9}\left[
\begin{array}{ccc}
\scriptstyle(1,1) & \scriptstyle(1,1) & \scriptstyle (1,1)_{a} \\
\scriptstyle(A^{1}) & \scriptstyle(A^{2}) & \scriptstyle(A)
\end{array}
\right] \renewcommand{\arraystretch}{0.9}\left[
\begin{array}{ccc}
\scriptstyle(1,1) & \scriptstyle(1,1) & \scriptstyle (1,1)_{a} \\
\scriptstyle (A^{3}) & \scriptstyle (A^{4}) & \scriptstyle (-A)
\end{array}
\right] \Bigl\}. \notag \\  \label{A9}
\end{align}
In the last SU(3)Clebsch-Gordan coefficients only the
antisymmetric irrep coupling is included and there is no summation
over irrep multiplicity. The capital Latin character indices $(A)$
denote the state label $(Z,I,M)$, $(-A)$ denotes $(-Z,I,-M)$ and
$(-1)^{A}=(-1)^{Z+M}$. The dependence on group irrep appear as an
overall factor because
\begin{equation}
\mathrm{Tr}\left\langle (\lambda ,\mu )\left|
J_{(A)}^{(1,1)}J_{(B)}^{(1,1)}\right| (\lambda ,\mu )
\right\rangle =(-1)^{A}%
\frac{1}{8}\mathrm{\dim }(\lambda ,\mu )C_{2}^{SU(3)}(\lambda ,\mu
)\delta _{(A),(-B)}\, ,  \label{G5}
\end{equation}
where $\mathrm{\dim }(\lambda ,\mu )=\frac{1}{2}(\lambda +1)(\mu
+1)(\lambda
+\mu +2)$ is a dimension of irrep. Above
$C_{2}^{SU(3)}(\lambda ,\mu )=\frac{1}{3}%
(\lambda ^{2}+\mu ^{2}+\lambda \mu +3\lambda +3\mu )$ is an
eigenvalue of the quadratic Casimir operator of SU(3):
\begin{equation}
\hat{C}_{2}^{SU(3)}(\lambda ,\mu
)=(-1)^{Z+M}J_{(Z,I,M)}^{(1,1)}J_{(-Z,I,-M)}^{(1,1)}.  \label{G6}
\end{equation}

The time component of the conserved topological current in the
Skyrme model represents the baryon number density which in terms of
the variables $\alpha ^{i}(x,t) $ takes the form:
\begin{align}
B^{0}(x) =&\frac{1}{24\pi ^{2}N}\text{\thinspace }\epsilon ^{0ikl}\text{%
\thinspace \textrm{Tr}\thinspace }\left( \partial _{i}U\right) U^\dagger%
\left( \partial _{k}U\right) U^\dagger\left( \partial _{l}U\right)
U^\dagger  \notag \\
=&\frac{(-1)^{A}}{2^{7}\sqrt{3}\pi ^{2}N}\dim (\lambda ,\mu
)C_{2}^{SU(3)}(\lambda ,\mu )\varepsilon ^{abc}\partial _{a}\alpha
^{i}C_{i}^{(A)}(\alpha )  \notag \\
&\times \partial _{b}\alpha ^{i^{\prime }}C_{i^{\prime
}}^{(A^{\prime })}(\alpha )\partial _{c}\alpha ^{i^{\prime \prime
}}C_{i^{\prime \prime }}^{(A^{\prime \prime })}(\alpha )\left[
\begin{array}{ccc}
\scriptstyle(1,1) & \scriptstyle(1,1) & \scriptstyle(1,1)_{a} \\
\scriptstyle(A^{\prime }) & \scriptstyle(A^{\prime \prime }) &
\scriptstyle(-A)
\end{array}
\right].  \label{G12}
\end{align}

For the classical chiral symmetric Skyrme model the
dependence on the
irrep is contained in the overall factor $N$. The normalization factor,
\begin{equation}
N=\frac{1}{4}\dim (\lambda ,\mu )C_{2}^{SU(3)}(\lambda ,\mu )\,,
  \label{G13}
\end{equation}
is chosen so as to that the smallest non trivial
baryon number equals unity: $B=\int \mathrm{d}%
^{3}xB^{0}(x)=1$. The dynamics of the system are independent of
the overall factor in Lagrangian. Therefore in the
classical case the Skyrme model defined in arbitrary irrep is
equivalent to the Skyrme model in the fundamental representation
$(1,0)$, for which $N=1$ .

The classical soliton solution of the hedgehog type for $(\lambda
,\mu )$ irrep of the SU(3) group may be expressed as direct sum of
hedgehog ans\"atze in SU(2) irreps \cite{Acus97}.
The SU(2) representations embedded in the $(\lambda
,\mu )$ irrep are defined by the canonical $SU(3)\supset SU(2)$
chain. The hedgehog generalization takes the form:
\begin{equation}
\exp i(\mathbf{\sigma }\cdot \hat{x})F(r)\rightarrow U_{0}\left( \hat{x}%
,F(r)\right) =\exp i2\left( J_{(0,1,\cdot )}^{(1,1)}\cdot \hat{x}\right)
F(r)=\sum_{z,j}^{\text{ }(\lambda ,\mu )}\oplus D^{j}(\varkappa )\, ,
\label{G14}
\end{equation}
where $\mathbf{\sigma }$ are Pauli matrices and $\hat{x}$ is
the unit vector. The
Euler angles of the SU(2) subgroup in terms of
polar angles $\varphi ,\theta $ and the chiral angle
function $F(r)$
are:
\begin{align}
\varkappa ^{1} =&\varphi -\arctan (\cos \theta \tan F(r))-\pi /2,
\notag \\
\varkappa ^{2} =&-2\arcsin (\sin \theta \sin F(r)),  \notag \\
\varkappa ^{3} =&-\varphi -\arctan (\cos \theta \tan F(r))+\pi /2.
\label{G15}
\end{align}
The normalization factor (\ref{G13}) ensures that the baryon number density
for the hedgehog skyrmion in a general irrep has the usual form:
\begin{align}
B^{0}(x) =& \, \frac{1}{24\pi ^{2}N}\text{\thinspace }\epsilon ^{0ikl}\text{%
\thinspace Tr\thinspace }\left( \partial _{i}U_{0}\right) U_{0}^\dagger%
\left( \partial _{k}U_{0}\right) U_{0}^\dagger\left( \partial
_{l}U_{0}\right) U_{0}^\dagger  \notag \\
=& -\frac{1}{2\pi ^{2}}\frac{\sin ^{2}F(r)}{r^{2}}F^{\prime }(r)\,.
\label{G16}
\end{align}

With the hedgehog ansatz (\ref{G14}), and after renormalization with the
factor (\ref {G13}), the Lagrangian density (\ref{G2}) reduces to
the following simple form:
\begin{align}
\mathcal{L}_{cl}[F(r)] =&-\mathcal{M}_{cl}(F(r))=-\bigg\{{\frac{f_{\pi }^{2}%
}{2}}\Bigl(F^{^{\prime }2}+{\frac{2}{r^{2}}}\sin ^{2}\!F\Bigr)
\notag
\\
&+{\frac{1}{8e^{2}}}{\frac{\sin ^{2}\!F}{r^{2}}}\Bigl(2F^{^{\prime }2}+{%
\frac{\sin ^{2}\!F}{r^{2}}}\Bigr)\bigg\}\,. \label{G17}
\end{align}
Variation of the classical hedgehog soliton mass leads to
standard differential equation for the chiral angle $F(r)$.

The SU(3) chiral symmetry breaking term of Lagrangian density is
defined here as:
\begin{equation}
\mathcal{L}_{SB}=-{\mathcal{M}}_{SB}=-\frac{1}{N}\text{\thinspace }\frac{%
f_{\pi }^{2}}{4}\left[ m_{0}^{2}\text{\textrm{Tr} }
\left\{ U + U^\dagger%
-2\mathds{1}\right\} -2m_{8}^{2}
\text{ Tr}\left\{ \left( U + U^\dagger%
\right) J_{(0,0,0)}^{(1,1)}\right\} \text{ }\right] .  \label{G7}
\end{equation}
This form is chosen so that it reduces to the mass term of the $\pi
,K,\eta $
mesons when the unitary field $U(\mathbf{x},t)=\exp [\frac{i}{f_{\pi }}%
\varphi _{k}\Lambda _{k}]$ is expanded around the classical vacuum
$U=\mathds{1}$ :
\begin{equation}
\mathcal{L}_{SB}=-\frac{1}{2}m_{\pi }^{2}(\varphi _{1}^{2}+\varphi
_{2}^{2}+\varphi _{3}^{2})-\frac{1}{2}m_{K}^{2}(\varphi
_{4}^{2}+\varphi _{5}^{2}+\varphi _{6}^{2}+\varphi
_{7}^{2})-\frac{1}{2}m_{\eta }^{2}\varphi _{8}^{2}+...\,\, .
\label{G8}
\end{equation}

For arbitrary irrep the coefficients in the symmetry breaking term can be
readily obtained as:
\begin{equation}
m_{0}^{2}=\frac{1}{3}\left( m_{\pi }^{2}+2m_{K}^{2}\right) ,\hspace{1cm}%
m_{8}^{2}=\frac{10}{3\sqrt{3}}\text{\thinspace }\frac{%
C_{2}^{SU(3)}(\lambda ,\mu )}{C_{3}^{SU(3)}(\lambda ,\mu )}\left( m_{\pi
}^{2}-m_{K}^{2}\right)\, ,  \label{G9}
\end{equation}
where
\begin{equation}
C_{3}^{SU(3)}(\lambda ,\mu )=\frac{1}{9}(\lambda -\mu )(2\lambda +\mu
+3)(2\mu +\lambda +3)\, ,  \label{G10}
\end{equation}
is the eigenvalue of the cubic Casimir operator of SU(3).

For the self adjoint irreps $\lambda =\mu $ the symmetry breaking part
of Lagrangian (\ref{G7}) is proportional to
$m_{0}^{2}=\frac{1}{4}m_{\pi }^{2}$ only. The Gell-Mann-Okubo mass
formula:
\begin{equation}
m_{\pi }^{2}+3m_{\eta }^{2}-4m_{K}^{2}=0\, ,  \label{G11}
\end{equation}
is satisfied in all but the self adjoint irreps.

\section{Quantization of the skyrmion}

The direct quantization of the Skyrme model even
in the case of SU(2)
leads to rather complicated equations \cite{Norvaisas}.
Here
the collective coordinates \cite{Adkins} for the unitary
field $U$ in $(\lambda ,\mu )$ irrep are employed for
the separation of the variables,
which depend on the temporal and spatial coordinates:
\begin{equation}
U(\hat{x},F(r),\mathbf{q}(t))=A(\mathbf{q}(t))U_{0}\left( \hat{x}%
,F(r)\right) A^\dagger(\mathbf{q}(t))\,.  \label{B1}
\end{equation}

Because of form of the ansatz $U_{0}$ (\ref{G14}), the unitary field
$U$ is invariant under right U(1) transformation of
the $A(\mathbf{q}(t))=$ $%
D^{(\lambda ,\mu )}(\mathbf{q}(t))$ matrix, defined as
\begin{equation}
A(\mathbf{q}(t))\rightarrow A(\mathbf{q}(t))\exp \beta
J_{(0,0,0)}^{(1,1)}\,. \label{B1a}
\end{equation}
Thus the seven-dimensional homogeneous space
SU(3)/U(1), which is specified by the seven real, independent
parameters $q^{k}(t)$, has
to be considered. The mathematical structure of the
Skyrme model and its quantization problems on the coset space
SU(3)/U(1) have been examined by several authors
\cite{Witten,Mazur}. The canonical quantization procedure
for the SU(3)
Skyrme model in the fundamental representation has been
considered by Fujii {\it et al.}
\cite{Fujii88}.
Here the attention is on the representation
dependence of the model. The Lagrangian (\ref{G2}) is considered
quantum mechanically
\textit{ab initio}. The generalized coordinates $q^{k}(t)$
and velocities $%
(\mathrm{d}/\mathrm{d}t)q^{k}(t)=\dot{q}^{k}(t)$ satisfy the
commutation relations:
\begin{equation}
\left[ \dot{q}^{k},q^{l}\right] =-if^{kl}(q)\,,  \label{B2}
\end{equation}
where $f^{kl}(q)$ are functions only of $q^{k}$, and the form of which
will be
determined below. The commutation relation between a velocity component $%
\dot{q}^{k}$ and arbitrary function $G(q)$ is given by
\begin{equation}
\left[ \dot{q}^{k},G(q)\right] =-i\sum_{r}f^{kr}(q)\partial
_{r}G(q)\,. \label{B3}
\end{equation}
For the time derivative the usual Weyl ordering is adopted:
\begin{equation}
\partial _{0}G(q)=\frac{1}{2}\left\{ \dot{q}^{k},\frac{\partial }{\partial
q^{k}}G(q)\right\}\, .  \label{B4}
\end{equation}
The operator ordering is fixed by the form of the Lagrangian
(\ref{G2}), without further ordering ambiguity.

The ansatz (\ref{B1}) is then substituted
in the Skyrme Lagrangian (\ref{G2}) followed by an
integration over the spatial coordinates. The
Lagrangian is then obtained
in terms of collective coordinates and velocities.
For the derivation of the canonical momenta it
is sufficient to restrict the
consideration to terms of second order
in the velocities here (the terms of first order vanish).
This leads to:
\begin{eqnarray}
&&L_{Sk}\approx \, -
\int drr^{2}\Bigl[\sum_{M}(-1)^{M}\left\{ \dot{q}%
^{i},C_{i}^{\prime (0,1,M)}(q)\right\} \left\{ \dot{q}^{i^{\prime
}},C_{i^{\prime }}^{\prime (0,1,-M)}(q)\right\} \notag \\
&&\phantom{IIIII}\times \frac{\pi }{3}\sin ^{2}F\left( f_{\pi
}^{2}+\frac{1}{\mathrm{e}^{2}}\left( F^{\prime
2}+\frac{1}{r^{2}}\sin
^{2}F\right) \right) \notag \\
&& \hspace{-0.5cm} +\sum_{Z,M}(-1)^{Z+M}
\left\{ \dot{q}^{i},C_{i}^{\prime (Z,\frac{1}{2}%
,M)}(q)\right\} \left\{ \dot{q}^{i^{\prime }},C_{i^{\prime }}^{\prime (-Z,%
\frac{1}{2},-M)}(q)\right\} \notag \\
&&\phantom{IIIII}\times \frac{\pi }{4}\left( 1-\cos
F\right) \left( f_{\pi }^{2}+\frac{1}{4\mathrm{e}^{2}}\left( F^{\prime 2}+%
\frac{2}{r^{2}}\sin ^{2}F\right) \right) \Bigl] \notag
\\
&&\approx \,  \frac{1}{2}\,
\dot{q}^{\alpha }g_{\alpha \beta }(q,F)\dot{q}%
^{\beta }+\left[ (\dot{q})^{0}\mathrm{-order \,\, term}\right]  \notag \\
&&\approx\, \frac{1}{8}\left\{ \dot{q}^{\alpha
},C_{\alpha }^{\prime (A)}(q)\right\} E_{(A)(B)}(F)\left\{
\dot{q}^{\beta },C_{\beta }^{\prime (B)}(q)\right\} +\left[
(\dot{q})^{0}\mathrm{-order \,\, term}\right] \label{B5}.
\end{eqnarray}

The Lagrangian (\ref{B5}) is normalized by the
factor (\ref{G13}). The metric
tensor takes the form
\begin{equation}
g_{\alpha \beta }(q,F)=C_{\alpha }^{\prime (A)}(q)E_{(A)(B)}(F)C_{\beta
}^{\prime (B)}(q)\,.  \label{B6}
\end{equation}
where
\begin{equation}
E_{(Z,I,M)(Z^{\prime },I^{\prime },M^{\prime
})}(F)=-(-1)^{Z+M}a_{I}(F)\delta _{Z,-Z^{\prime }}\delta
_{I,I^{\prime }}\delta _{M,-M^{\prime }}\,.  \label{B7}
\end{equation}
Here the soliton moments of inertia are given as
integrals over the dimensionless variable $\tilde{r}=\mathrm{%
e}f_{\pi }r$:
\begin{subequations}
\begin{align}
a_{0}(F)& =0,  \\
a_{\frac{1}{2}}(F)& =\frac{1}{\mathrm{e}^{3}f_{\pi }}\tilde{a}_{\frac{1}{2}%
}(F)=\frac{1}{\mathrm{e}^{3}f_{\pi }}2\pi \int
d\tilde{r}\tilde{r}^{2}\left(
1-\cos F\right) \left[ 1+\frac{1}{4}F^{\prime 2}+\frac{1}{2 \tilde{r}%
^{2}}\sin ^{2}F\right] , \label{B81} \\
a_{1}(F)& =\frac{1}{\mathrm{e}^{3}f_{\pi }}\tilde{a}_{1}(F)=\frac{1}{\mathrm{%
e}^{3}f_{\pi }}\frac{8\pi }{3}\int d\tilde{r}\tilde{r}^{2}\sin
^{2}F\left[ 1+F^{\prime 2}+\frac{1}{r^{2}}\sin ^{2}F\right]\, .
\label{B8}
\end{align}
\end{subequations}

The canonical momentum, which is conjugate to ${q}^{\beta }$, is
defined as
\begin{equation}
p_{\beta }^{(0)}=\frac{\partial L_{Sk}}{\partial
\dot{q}^{\beta }}=\frac{1}{2%
}\left\{ \dot{q}^{\alpha },g_{\alpha \beta }\right\}\, .  \label{B9}
\end{equation}
The canonical commutation relations
\begin{align}
\left[ q^{\alpha },q^{\beta }\right] =&\left[ p_{\alpha
}^{(0)},p_{\beta }^{(0)}\right]=0\,, \notag
\\
\left[ p_{\beta }^{(0)},q^{\alpha }\right] =&-i\delta _{\alpha
\beta }\,, \label{B10}
\end{align}
then yield the following explicit form for the functions $f^{\alpha \beta
}(q)$:
\begin{equation}
f^{\alpha \beta }(q)=\left( g_{\alpha \beta }\right) ^{-1}=C_{(\bar{A}%
)}^{^{\prime }\alpha
}(q)E^{(\bar{A})(\bar{B})}(F)C_{(\bar{B})}^{^{\prime }\beta }(q)\,,
\label{B11}
\end{equation}
where
\begin{equation}
\ E^{(\overline{Z,I,M})(\overline{Z^{\prime },I^{\prime },M^{\prime }}%
)}(F)=-(-1)^{Z+M}\frac{1}{a_{I}(F)}\delta _{Z,-Z^{\prime }}\delta
_{I,I^{\prime }}\delta _{M,-M^{\prime }}\, .  \label{B12}
\end{equation}
Note that here $E^{(0)(0)}(F)$ is left undefined. The
summation over the indices $(%
\bar{A})$ denotes summation over the basis states $(Z,I,M)$ of irrep
$(1,1)$, excluding the state $(0,0,0)$. It proves convenient
to introduce the
reciprocal function matrix $C_{(%
\bar{A})}^{^{\prime }\alpha }(q)$, the properties of which
are described
in the Appendix. The commutation relations of
the momenta (\ref{B10})
ensure the choice
of parameters\ $q^{\alpha }$ on the manifold SU(3)/U(1) (see \cite{Fujii87}%
). Here there is no need for explicit parameterization of $q^{\alpha
}$.

After determination of function $f^{\alpha \beta }(q)$ the
following explicit expression $A^\dagger\dot A$ obtains:
\begin{align}
A^\dagger\dot{{A}}& =\frac{1}{2}D^{(\lambda ,\mu
)}(-q)\left\{ \dot{q}^{\alpha },\partial _{\alpha }D^{(\lambda
,\mu
)}(q)\right\} \notag \\
& =\frac{1}{2}\left\{ \dot{q}^{\alpha },C_{\alpha }^{\prime
(A)}(q)\right\}
\left\langle \hspace{0.2cm}\left| \hspace{0.1cm}J_{(A)}^{(1,1)}\hspace{0.05cm%
}\right| \hspace{0.2cm}\right\rangle  \notag \\
&\phantom{W} -\frac{1}{2}iE^{(\bar{A})(\bar{B})}(F)C_{(\bar{B}%
)}^{^{\prime }\beta }(q)C_{\beta }^{^{\prime }(0)}(q)\left(
\left\langle
\hspace{0.2cm}\left| \hspace{0.1cm}J_{(0)}^{(1,1)}J_{(\bar{A})}^{(1,1)}%
\hspace{0.05cm}\right| \hspace{0.2cm}\right\rangle +\left\langle \hspace{%
0.2cm}\left| \hspace{0.1cm}J_{(\bar{A})}^{(1,1)}J_{(0)}^{(1,1)}\hspace{0.05cm%
}\right| \hspace{0.2cm}\right\rangle \right)  \notag \\
&\phantom{W} -\frac{3}{8}iE^{(\bar{A})(\bar{B})}(F)C_{(\bar{A}%
)}^{^{\prime }\alpha }(q)C_{\alpha }^{^{\prime }(0)}(q)C_{(\bar{B}%
)}^{^{\prime }\beta }(q)C_{\beta }^{^{\prime
}(0)}(q)\sum_{z,j}^{(\lambda
,\mu )}\oplus \,\,y^{2}\,\mathds{1}_{z,j} \notag \\
&\phantom{W} +\frac{i}{2
a_{\frac{1}{2}}(F)}\,C_{2}^{SU(3)}(\lambda
,\mu )\mathds{1}+i\sum_{z,j}^{(\lambda ,\mu )}\oplus \left( \frac{%
C^{SU(2)}(j)}{2
a_{1}(F)}-\frac{C^{SU(2)}(j)+\frac{3}{4}\,y^{2}}{2
a_{\frac{1}{2}}(F)}\right) \mathds{1}_{z,j}\, .  \label{B13}
\end{align}
Here $\mathds{1}$ is the unit matrix in the $(\lambda ,\mu )$
irrep of SU(3) and $\mathds{1}_{z,j}$ are unit matrices in the SU(2)
irreps. Note that the inverse of the rotation
represented by $D^{(\lambda,\mu)}(q)$ is denoted
$D^{(\lambda,\mu)}(-q)$.

The field expression (\ref{B1}) is substituted
in the  Lagrangian density (\ref{G2}) in order to obtain
the explicit expression in terms of collective coordinates and
space coordinates. Some  expressions with SU(3) group
generators, that are useful for this purpose,
are presented in Appendix. After some lengthy manipulation
the complete expression of the Skyrme model Lagrangian
density is obtained as:
\begin{align}
{\mathcal{L}}_{Sk}& =\frac{1}{4}\dim (\lambda ,\mu
)C_{2}^{SU(3)}(\lambda
,\mu )\times   \notag \\
& \times \Bigl\{-\frac{(1-\cos F)}{16}\left[ f_{\pi }^{2}+\frac{1}{4e^{2}}%
\left( F^{\prime 2}+\frac{2}{r^{2}}\sin ^{2}F\right) \right]   \notag \\
& \phantom{WWWWWWW}\times \sum_{Z,M} \, (-1)^{Z+M}\left\{
\dot{q}^{\alpha },C_{\alpha }^{\prime
(Z,\frac{1}{2},M)}(q)\right\} \left\{ \dot{q}^{\beta },C_{\beta }^{\prime (Z,%
\frac{1}{2},M)}(q)\right\}  \notag \\
& \phantom{Wii}-\frac{\sin ^{2}F}{8}\left[ f_{\pi
}^{2}+\frac{1}{e^{2}}\left(
F^{\prime 2}+\frac{1}{r^{2}}\sin ^{2}F\right) \right]   \notag \\
& \phantom{WWWWWWW}\times \sum_{Z,M} \, \Bigl[(-1)^{M}\Bigl\{
\dot{q}^{\alpha },C_{\alpha }^{\prime (0,1,M)}(q)\Bigl\} \left\{
\dot{q}^{\alpha ^{\prime }},C_{\alpha
^{\prime }}^{\prime (0,1,M)}(q)\right\} \notag \\
& \phantom{WWWWWWW}-\Bigl( \Bigl\{ \dot{q}^{\alpha
},C_{\alpha}^{\prime (0,1,\cdot )}(q)\Bigl\} \, \cdot \,
\hat{x}\Bigl) \left( \left\{ \dot{q}^{\alpha ^{\prime }},C_{\alpha
^{\prime }}^{\prime (0,1,\cdot )}(q)\right\} \cdot
\hat{x}\right) \Bigl]  \notag \\
& \phantom{Wii}-{\mathcal{M}}_{cl}-
\Delta {\mathcal{M}}_{1}-\Delta {\mathcal{M}%
}_{2}-\Delta {\mathcal{M}}_{3}-\Delta {\mathcal{M}}^{\prime
}(q)\Bigl\}\, . \label{B14}
\end{align}
Here the following notation has been introduced:

\begin{subequations}
\begin{align}
\begin{split}
\Delta {\mathcal{M}}_{1}(F)& =-\frac{\sin ^{2}F}{30 a_{1}^{2}(F)}
\Bigl[f_{\pi }^{2}\left( 12\sin ^{2}F\cdot
C_{2}^{SU(3)}(\lambda ,\mu )-16\sin ^{2}F+15\right) \hspace{5cm} \\
& \phantom{W}+\frac{1}{2e^{2}}\Bigl( 2F^{\prime 2}\left( 12\cos
^{2}F\cdot
C_{2}^{SU(3)}(\lambda ,\mu )+16\sin ^{2}F-1\right) \\
&  \phantom{W+\frac{1}{2e^{2}}\Bigl(}+\frac{\sin ^{2}F}{r^{2}}%
\left( 6C_{2}^{SU(3)}(\lambda ,\mu )+7\Bigl) \right)
\Bigl]\,,\hspace{5cm} \label{kvmasesa}
\end{split}
\\
\begin{split}
\Delta {\mathcal{M}}_{2}(F)& =
-\frac{\left( 1-\cos F\right) }{20 a_{%
\frac{1}{2}}^{2}(F)}\Bigl[f_{\pi }^{2}\left( 6\left( 1-\cos
F\right) \cdot
C_{2}^{SU(3)}(\lambda ,\mu )+3\cos F+2\right)  \\
& \phantom{W}+\frac{1}{4e^{2}}\left( F^{\prime 2}\left( 6\left(
1+\cos F\right) \cdot C_{2}^{SU(3)}(\lambda ,\mu )-3\cos
F+2\right) +10\frac{\sin ^{2}F}{r^{2}}\right) \Bigl]\,,
\label{kvmasesb}
\end{split}
\\
\begin{split}
\Delta {\mathcal{M}}_{3}(F)& =-\frac{\sin ^{2}F}{30 a_{1}(F)\,
a_{\frac{1}{2}}(F)}\Bigl[f_{\pi }^{2}\left(
12\left( 1-\cos
F\right) \cdot C_{2}^{SU(3)}(\lambda ,\mu )+16\cos F-1\right)  \\
& \phantom{W}+\frac{1}{2e^{2}}\left( F^{\prime 2}\left( 4\cos
F\cdot \left(
3C_{2}^{SU(3)}(\lambda ,\mu )-4\right) +15\right)
+15\frac{\sin ^{2}F}{r^{2}}%
\right) \Bigl]\,, \hspace{5cm} \label{kvmasesc}
\end{split}
\\
\begin{split}
\Delta {\mathcal{M}}^{\prime }(F,q)& =-\frac{3(1-\cos F)}{16a_{\frac{1}{2}%
}^{2}(F)}\left[ f_{\pi }^{2}+\frac{1}{4e^{2}}\left( F^{\prime 2}+\frac{2}{%
r^{2}}\sin ^{2}F\right) \right] \hspace{7cm} \\
& \phantom{W}\times \left( (-1)^{\overset{\phantom{i}=}{A}}C_{(%
\overset{\phantom{i}=}{A})}^{^{\prime }\alpha }(q)C_{\alpha
}^{^{\prime }(0)}(q)C_{(-\overset{\phantom{i}=}{A})}^{^{\prime
}\beta }(q)C_{\beta }^{^{\prime }(0)}(q)+4\right)\, .
\label{kvmasesd}
\end{split}
\end{align}
\end{subequations}
[The ``4'' in the last bracket on the last row is missing
in the corresponding
expression in ref.\cite{Fujii88}, the consequence of which
is the appearance of a spurious term $-3/8a_{1/2\, (F)}$
in eq. (69b) of
that paper (there are some minor misprints in that equation as well)].
The notation ($\overset{\phantom{i}=}{A}$) indicates
that only the states for which
$I=\frac{1}{2}$ and $Z=\pm \frac{1}{2}$ are included.
The $\Delta
{\mathcal{M}}_{k}(F)$ terms may be interpreted as quantum mass
corrections to the Lagrangian density. The $\Delta
{\mathcal{M}}^{\prime }(F,q)$ term depends on the quantum variables
$q^{i}$ and is an operator on the configuration space.

The integration (\ref{B15}) over the space variables and normalization by
factor (\ref{G13}) gives the Lagrangian
\begin{align}
L_{Sk} &=\int {\mathcal{L}}_{Sk}\mathrm{d}^{3}x=\frac{1}{8}\left\{ \dot{q}%
^{i},C_{i}^{\prime (\bar{A})}(q)\right\} E_{(\bar{A})(\bar{B})}
\left\{ \dot{q%
}^{i^{\prime }},C_{i^{\prime }}^{\prime (\bar{B})}(q)\right\}
\notag
\\
& \phantom{W}-M_{cl}-\Delta M_{1}-\Delta M_{2}-\Delta M_{3}-\Delta
M^{\prime }(q)
\notag \\
&=-\frac{1}{8a_{\frac{1}{2}}(F)}(-1)^{\bar{A}}\left\{ \dot{q}%
^{i},C_{i}^{\prime (\bar{A})}(q)\right\} \left\{
\dot{q}^{i^{\prime
}},C_{i^{\prime }}^{\prime (-\bar{A})}(q)\right\} \notag \\
& \phantom{W}-\frac{1}{8}\left( \frac{1%
}{a_{1}(F)}-\frac{1}{a_{\frac{1}{2}}(F)}\right) (-1)^{M}\left\{
\dot{q}^{i},C_{i}^{\prime (0,1,M)}(q)\right\} \left\{
\dot{q}^{i^{\prime }},C_{i^{\prime }}^{\prime (0,1,-M)}(q)\right\}
\notag \\
& \phantom{W}-M_{cl}-\Delta M_{1}-\Delta M_{2}-\Delta M_{3}-\Delta
M^{\prime }(q)\,. \label{B15}
\end{align}
Here $M_{cl}=\frac{f_{\pi}}{e}\tilde{M}_{cl}=\int
\mathrm{d}^{3}x\mathcal{M}_{cl}(F)$, $\Delta
M_{k}=e^{3}f_{\pi}\Delta\tilde{M}_{k}=\int \mathrm{d}^{3}x\Delta
\mathcal{M}_{k}(F)$ and $\Delta M^{\prime }(q)=\int
\mathrm{d}^{3}x \Delta \mathcal{M}^{\prime }(q)$, where
$\tilde{M}_{cl}$ and $\tilde{M}_{k}$ are integrals over the
dimensionless variable.

\section{Structure of the Lagrangian and the Hamiltonian}

\bigskip

The Wess-Zumino-Witten (WZW) action is given as an integral over
the five dimensional manifold $M^{5}$, the boundary of which is
the compactified spacetime: $\partial M^{5}=M^{4}=S^{3}\times
S^{1}$. This term is necessary to account for the anomalies in QCD
\cite{Witten}. The standard form for this term is:
\begin{equation}
S_{WZ}(U)=-\frac{iN_{c}}{240\pi ^{2}N^{\prime }}\int\limits_{M^{5}}d^{5}x%
\text{\thinspace }\epsilon ^{\mu \nu \lambda \rho \sigma }\text{\thinspace
Tr\thinspace }R_{\mu }R_{\nu }R_{\lambda }R_{\rho }R_{\sigma }\, ,
 \label{C1}
\end{equation}
where $N_{c}$ is the number of colors and $N^{\prime }$ is a
normalization factor. The derivation of the
contribution of the Wess-Zumino-Witten term to the effective
Lagrangian in the framework of collective coordinate formalism is
given in ref. \cite{Balachan}. By application of Stoke's theorem
it takes the following form in a general dimension:
\begin{eqnarray}
&&L_{WZ}(q,\dot{q}) =-\frac{iN_{c}}{24\pi ^{2}N^{\prime }}%
\int\limits_{M^{3}}d^{3}x\text{\thinspace }\epsilon
^{mjk}\text{\thinspace
Tr\thinspace }\left[ \left( \partial _{m}U_{0}\right) U_{0}^\dagger
\left( \partial _{j}U_{0}\right) U_{0}^\dagger\left( \partial
_{k}U_{0}\right) U_{0}^\dagger J_{(0,0,0)}^{(1,1)}\right]
\nonumber\\
&&\times \frac{1}{2}\left\{ \dot{q}^{\alpha },C_{\alpha
}^{\prime
(0)}(q)\right\}   \nonumber \\
&&=-\frac{iN_{c}}{2\sqrt{3}\pi ^{2}N^{\prime }}\int\limits_{M^{3}}d^{3}x%
\frac{\sin ^{2}F(r)}{r^{2}}F^{\prime }(r)\sum_{z,j}^{(\lambda ,\mu
)}yj(j+1)(2j+1)
\frac{1}{2}\text{\thinspace }\left\{
\dot{q}^{\alpha
},C_{\alpha }^{\prime (0)}(q)\right\}   \nonumber \\
&&=-\lambda ^{\prime }\frac{i}{2}\left\{ \dot{q}^{\alpha
},C_{\alpha }^{\prime (0)}(q)\right\} \,.  \label{C2}
\end{eqnarray}
Here
\begin{equation}
\lambda ^{\prime }=\frac{\sqrt{3}N_{c}B}{40N^{\prime }}\text{%
\thinspace }\dim (\lambda ,\mu )C_{3}^{SU(3)}(\lambda ,\mu )\,.
  \label{C3}
\end{equation}
The coefficient $\lambda ^{\prime }$ depends on the
representation $(\lambda ,\mu )$ . For all self adjoint irreps
$\lambda =\mu $ the WZW term vanishes. Following Witten
\cite{Witten} the normalization factor is chosen to be $N^{\prime
}=\dim (\lambda ,\mu
)C_{3}^{SU(3)}(\lambda ,\mu )/20$ so that
 $ \lambda ^{\prime }=N_c B
/2\sqrt{3}$\thinspace. In the fundamental representation
$N^{\prime }=1$. Here the coefficient $\lambda
^{\prime }$ only serves to constrain the states of the system.
Because the cubic Casimir
operator $C_3^{SU(3)}$ (\ref{G10}) vanishes in the self-adjoint
representations, it follows that
the WZW term (\ref{C2}) also vanishes in those
representations.

The Lagrangian of the system, with inclusion of the
WZW term is:
\begin{equation}
L^{\prime }=L_{Sk}+L_{WZ}\,  .  \label{C4}
\end{equation}
There are seven
collective coordinates.
The momenta $p_{\alpha }$ that are canonically
conjugate to $q^{\alpha }$ are defined
as
\begin{equation}
p_{\alpha }=\frac{\partial L^{\prime }}
{\partial \dot{q}^{\alpha }}=\frac{1}{%
2}\left\{ \dot{q}^{\beta },g_{\beta \alpha })\right\} -i\lambda
^{\prime }C_{\alpha }^{^{\prime }(0)}(q)\,.  \label{C5}
\end{equation}
These satisfy the canonical commutation relations (\ref{B10}). The
WZW term may be considered as an
external potential in the system \cite{Mazur}. The seven right
transformation generators may be defined as:
\begin{equation}
\hat{R}_{(\bar{A})}=\frac{i}{2}\left\{ p_{\alpha }+\lambda ^{\prime
}iC_{\alpha }^{\prime (0)}(q),C_{(\bar{A})}^{\prime \alpha }(q)\right\} =%
\frac{i}{2}\left\{ \dot{q}^{\beta },C_{\beta }^{\prime (\bar{B})}(q)\right\}
E_{(\bar{B})(\bar{A})}\, .  \label{C6}
\end{equation}
The commutation rules for the generators (\ref{C6}) and
their action on the $%
D^{(\lambda ,\mu )}$ matrices are given in Appendix. It is
convenient to define
an eighth
transformation generator formally as \cite{Fujii88}:
\begin{equation}
\hat{R}_{(0)}=-\lambda ^{\prime }\,.  \label{C8}
\end{equation}
The SU(2) subalgebra of the operators
$\hat{R}_{(0,1,M)}$ satisfies the standard SU(2)
commutation conditions and may be interpreted as spin generators
(Appendix). The eight left transformation generators may be
defined as:
\begin{equation}
\hat{L}_{(B)}=\frac{1}{2}\left\{ \hat{R}_{(A)},D_{(A)(B)}^{(1,1)}(-q)\right%
\} .  \label{C9}
\end{equation}
The transformation properties and commutation relations for
the left
and right transformation generators are given in Appendix.

The effective Lagrangian, which includes the WZW term takes the
form:
\begin{align}
L_{eff} &=\frac{1}{2a_{\frac{1}{2}}(F)}(-1)^{\bar{A}}\hat{R}_{(\bar{A})}%
\hat{R}_{(-\bar{A})}+\left( \frac{1}{2a_{1}(F)}-\frac{1}{2a_{\frac{1}{2}}(F)}%
\right) \left( \hat{R}_{(0,1,\cdot )}\cdot \hat{R}_{(0,1,\cdot
)}\right)    \notag \\
& \phantom{W}-\lambda ^{\prime }\frac{i}{2}\left\{ \dot{q}^{\alpha
},C_{\alpha }^{\prime (0)}(q)\right\}-M_{cl}-\Delta M_{1}-\Delta
M_{2}-\Delta M_{3}-\Delta M^{\prime }(q)
\notag \\
&=\frac{1}{2a_{\frac{1}{2}}(F)}\left( (-1)^{A}\hat{L}_{(A)}\hat{L}%
_{(-A)}-\lambda ^{\prime 2}\right) +\left( \frac{1}{2a_{1}(F)}-\frac{1}{2a_{%
\frac{1}{2}}(F)}\right) \left( \hat{R}_{(0,1,\cdot )}\cdot \hat{R}%
_{(0,1,\cdot )}\right)   \notag \\
& \phantom{W}-\lambda ^{\prime }\frac{i}{2}\left\{ \dot{q}^{\alpha
},C_{\alpha }^{\prime (0)}(q)\right\} -\Delta M_{1}-\Delta
M_{2}-\Delta M_{3}-M_{cl}\,. \label{C10}
\end{align}
Note that the $\Delta {M}^{\prime }(q)$ term which depends on
quantum variables due to introducing of left translation
generators (see (\ref{lefttrans})) in the Lagrangian expression
(\ref{C10}) vanishes.

For the purpose of obtaining Euler-Lagrange equations that are
consistent with the canonical equation of motion of the
Hamiltonian, the general method of quantization on a curved space
developed by Sugano \textit{et al.} \cite{Sugano} is employed, in
which the following auxiliary function is introduced:
\begin{align}
Z(q) &=-\frac{1}{16}f^{ab}f^{cd}f^{ek}\left( \partial
_{a}g_{cd}\right) \left( \partial _{b}g_{ek}\right)
-\frac{1}{4}\partial _{a}\left( f^{ab}f^{cd}\partial
_{b}g_{cd}\right) -\frac{1}{4}\partial _{a}\partial
_{b}f^{ab}  \notag \\
&=-\frac{1}{4}\partial _{b}C_{(\bar{A})}^{\prime a}(q)E^{(\bar{A})(\bar{B}%
)}\partial _{a}C_{(\bar{B})}^{\prime b}(q)  \notag \\
& \phantom{W}+\frac{3}
{16a_{\frac{1}{2}}(F)}\left( (-1)^{\overset{\phantom{i}=}{A}%
}C_{a}^{^{\prime }(0)}(q)C_{(\overset{\phantom{i}=}{A})}^{^{\prime
}a}(q)C_{(-\overset{\phantom{i}=}{A})}^{^{\prime
}b}(q)C_{b}^{\prime (0)}(q)+4\right) .  \label{C11}
\end{align}
With this the covariant kinetic term may be defined as:
\begin{align}
2K &=\frac{1}{2}\left\{ p_{\alpha }+i\lambda C_{\alpha }^{\prime (0)}(q),%
\dot{q}^{\alpha }\right\} -Z(q)  \notag \\
&=\frac{1}{a_{\frac{1}{2}}(F)}\left( (-1)^{A}\hat{L}_{(A)}\hat{L}%
_{(-A)}-\lambda ^{\prime 2}\right) +\left( \frac{1}{a_{1}(F)}-\frac{1}{a_{\frac{1}{2%
}}(F)}\right) \left( \hat{R}_{(0,1,\cdot )}\cdot
\hat{R}_{(0,1,\cdot )}\right)\, . \notag \\
\label{C12}
\end{align}

According to the prescription
\cite{Sugano} the effective Hamiltonian (with the constraint
(\ref{C8})) is constructed in the
standard form as:
\begin{equation}
H=\frac{1}{2}\{p_{\alpha },\dot{q}^{\alpha
}\}-L_{eff}-Z(q)=K+\Delta M_{1}+\Delta M_{2}+\Delta
M_{3}+M_{cl}\,. \label{C13}
\end{equation}

Upon renormalization the Lagrangian density (\ref{B14}) may be
reexpressed in terms of left and right transformation generators.
The effective Hamiltonian density without the
symmetry breaking term
in turn takes the form:
\begin{align}
{\mathcal{H}}_{Sk} &=\frac{(1-\cos F)}{4
a_{\frac{1}{2}}^{2}(F)}\left[ f_{\pi }^{2}+\frac{1}{4e^{2}}\left(
F^{\prime 2}+\frac{2}{r^{2}}\sin
^{2}F\right) \right] \notag \\
& \phantom{WWWWWWWWWWW} \times \left[ (-1)^{A}\hat{L}_{(A)}
\hat{L}_{(-A)}-\left( \hat{R%
}_{(0,1,\cdot )}\cdot \hat{R}_{(0,1,\cdot )}\right) -\lambda ^{\prime 2}%
\right]   \notag \\
&\phantom{W}+ \frac{\sin ^{2}F}{2a_{1}^{2}(F)}\left[ f_{\pi }^{2}+\frac{%
1}{e^{2}}\left( F^{\prime 2}+\frac{1}{r^{2}}\sin ^{2}F\right)
\right] \notag \\
& \phantom{WWWWWWWWWWW} \times \left[ \left( \hat{R}_{(0,1,\cdot
)}\cdot \hat{R}_{(0,1,\cdot )}\right) -\left( \hat{R}_{(0,1,\cdot
)}\cdot \hat{x}\right) \left( \hat{R}_{(0,1,\cdot
)}\cdot \hat{x}\right) \right]   \notag \\
&\phantom{W}+\Delta {\mathcal{M}}_{1}+\Delta {\mathcal{M}}_{2}+\Delta {%
\mathcal{M}}_{3}+{\mathcal{M}}_{cl}\,.  \label{Htank}
\end{align}

The
products of spin operators $\hat{R}_{(0,1,M)}$ may be
separated into scalar and tensorial terms as:
\begin{align}
& \left( \hat{R}_{(0,1,\cdot )}\cdot \hat{R}_{(0,1,\cdot )}\right)
-\left( \hat{R}_{(0,1,\cdot )}\cdot \hat{x}\right) \left(
\hat{R}_{(0,1,\cdot )}\cdot \hat{x}\right) = \notag  \\
& = \frac{2}{3}\left( \hat{R}_{(0,1,\cdot )}\cdot
\hat{R}_{(0,1,\cdot )}\right)-\frac{4\pi }{3}Y_{2,M+M^{\prime
}}^{\ast }(\vartheta ,\varphi )\left[
\begin{array}{ccc}
\scriptstyle 1 & \scriptstyle 1 & \scriptstyle 2 \\
\scriptstyle M & \scriptstyle M^{\prime } & \scriptstyle
M+M^{\prime }
\end{array}
\right] \hat{R}_{(0,1,M)}\hat{R}_{(0,1,M^{\prime })}\, , \label{C15}
\end{align}
where $Y_{l,M}(\vartheta ,\varphi )$ is a spherical harmonic and
the factor in the square brackets on the right-hand side is
an SU(2) Clebsch-Gordan coefficient.

The covariant kinetic term (\ref{C12}) is a differential operator
constructed from SU(3) left and SU(2) right transformation
generators. The eigenstates of the Hamiltonian (\ref{C13}) are:
\begin{equation}
\genfrac{|}{\rangle}{0pt}{}{\scriptstyle (\Lambda
,M)}{\scriptstyle Y,T,M_{T};Y^{\prime },S,M_{S}} =\sqrt{\dim
(\Lambda ,M)}D_{(Y,T,M_{T})(Y^{\prime },S,M_{S})}^{\ast (\Lambda
,M)}(q)\left| 0\right\rangle\, . \label{C16}
\end{equation}
Here the quantity $D$ on the right-hand side is the complex
conjugate Wigner matrix elements of $(\Lambda ,M)$ irrep of
SU(3) in terms of quantum variables $q^{k}$. The topology of
the eigenstates can be nontrivial and the quantum states contain
an eighth ''unphysical'' quantum variable $q^{0}$.

The matrix elements of the Hamiltonian density (\ref{Htank}) for
states with spin $S>\frac{1}{2}$ are not spherical and those
states consequently have quadrupole moments. In the
case $S=\frac{1}{2}$ the matrix element of \ the second rank
operator in right hand side of (\ref{C15}) vanishes.

\section{The symmetry breaking mass term}

The
chiral symmetry breaking mass term for the SU(3) soliton
was defined (\ref{G7}).
With  the
ansatz (\ref{B1}) in (\ref{G7}) the symmetry breaking
density operator for the general irrep $(\lambda,\mu)$ obtains as:
\begin{align}
\mathcal{L}_{SB}=-{\mathcal{M}}_{SB}=& -\frac{1}{N}\text{\thinspace }\frac{%
f_{\pi }^{2}}{4}\Bigl[ m_{0}^{2}\text{ Tr }\left\{ U_{0}{+}
U_{0}^\dagger
-2\mathds{1}\right\} \notag \\
& -2m_{8}^{2}\text{ Tr}\left\{ \left( U_{0}{+}
U_{0}^\dagger\right) J_{(0,0,0)}^{(1,1)}\right\}
\text{ }D_{(0)(0)}^{(1,1)}(-q)%
\Bigl]\,.   \label{S1}
\end{align}
The operator (\ref{S1}) contains the matrix elements $D^{(1,1)}$,
which depend on the quantum variables $q^{\alpha }$. In this form
this operator mixes the representations $(\Lambda ,M)$ of
the eigenstates of the Hamiltonian
\cite{Weigel}. The physical states of the system
with symmetry breaking term therefore in principle
have to be calculated by diagonalisation of the Hamiltonian.
Since the mass term is minor part of the Lagrangian
it may considered as a perturbation in the SU(3)
representation $(\Lambda,M)$.

For a given irrep $(\lambda,\mu)$, in which the Lagrangian is
defined, the symmetry breaking term depends on the chiral angle
$F(r)$ as:
\begin{align}
\text{Tr}\left\{ U_{0}{+}U_{0}^\dagger -2\mathds{1}\right\}& =
2\sum_{z,j}^{(\lambda ,\mu )}\left( \sum_{m=-j}^{j}\cos
2mF(r)\right)
-2\dim (\lambda ,\mu )  \notag \\
& =2\text{\thinspace }\frac{\sin (1+\lambda )F(r)+\sin (1+\mu
)F(r)-\sin (\lambda +\mu +2)F(r)}{2\sin F(r)-\sin 2F(r)} \notag \\
& \phantom{W} -2\dim (\lambda ,\mu )\, . \label{S2}
\end{align}
Further development of the expression (\ref{S1}) leads to:
\begin{align}
&\text{Tr\thinspace }\left\{ \left(
U_{0}{+}U_{0}^\dagger \right) J_{(0,0,0)}^{(1,1)}\right\}
=2\sum_{z,j}^{(\lambda ,\mu )}2\sqrt{3}\left[ \frac{1}{3}(\lambda -\mu )+z%
\right] \left( \sum_{m=-j}^{j}\cos 2mF(r)\right)   \notag \\
&=\frac{2\sqrt{3}}{2\sin F(r)-\sin 2F(r)} \notag \\
&\phantom{W}\times \Bigl\{\frac{1}{2}(1+\mu )\left( \sin (1+\mu
)F(r)-\sin (\lambda
+\mu +2)F(r)\right)   \notag \\
&\phantom{W}+\frac{1}{3}(\lambda -\mu )\left( \sin (1+\lambda
)F(r)+\sin
(1+\mu )F(r)-\sin (\lambda +\mu +2)F(r)\right)   \notag \\
&\phantom{W}+\frac{1}{2}(1+\lambda )\Bigl[\left( \sin F(r)-\sin
(2+\mu
)F(r)\right)\cos \lambda F(r) \notag \\
&\phantom{W}-\left( \cos F(r)-\cos (2+\mu )F(r)\right)
\sin \lambda F(r)
\Bigl]\Bigl\}\,. \label{S3}
\end{align}

For high irrep $(\lambda,\mu)$
the dependence of the symmetry breaking term on
the chiral angle $F(r)$ differs significantly from that
in the fundamental representation $\ (1,0)$. In that
representation the
symmetry breaking term takes the standard form:
\begin{equation}
{\mathcal{M}}_{SB}=f_{\pi }^{2}\text{ }(1-\cos F)\left[ m_{0}^{2}+\frac{1}{%
\sqrt{3}}\, m_{8}^{2}D_{(0)(0)}^{(1,1)}(-q)\right]\, .   \label{S4}
\end{equation}
In the case of $(2,0)$ representation the expression is:
\begin{equation}
{\mathcal{M}}_{SB}=\frac{1}{5}f_{\pi }^{2}\left[ (1-\cos F+2\sin
^{2}F)\, m_{0}^{2}-(1-\cos F-4\sin ^{2}F)\frac{m_{8}^{2}}{\sqrt{3}}\text{ }%
D_{(0)(0)}^{(1,1)}(-q)\right]\,.
\end{equation}
Note that in both cases the asymptotical behavior at large
distance of the symmetry
breaking terms are different.

\section{Discussion}

Above the SU(3) Skyrme model was quantized canonically in the
framework of the collective coordinate formalism in for
representations of arbitrary dimension. This lead to the
complete quantum mechanical structure of the model on the
homogeneous space SU(3)/U(1). The results extend those
obtained earlier in the fundamental representation for
SU(2) and SU(3) \cite{Fujii87, Fujii88} and those obtained
in general representations of SU(2)
\cite{Norvaisas, Acus97, Acus98}. The explicit representation
dependence of the quantum corrections
to the Skyrme
model Lagrangian was derived. This dependence is
nontrivial, especially for the Wess-Zumino-Witten and the symmetry
breaking terms.
The operators that form the Hamiltonian were shown to have well
defined group-theoretical properties.

The choice of the irrep that
is used for the unitary field depends on the phenomenological
aspects of the physical system to which the model is applied.
Formally the variation of the irrep can by interpreted as
modification of the Skyrme model. The representation dependence of
the Wess-Zumino-Witten term was shown to
absorbable into an normalization
factor, with exception of the self adjoint irreps
in which this term vanishes. The symmetry breaking term has
different functional dependence on chiral angle $F(r)$ in
different irreps. In case of self adjoint representations the
symmetry breaking term, which is proportional to $m_{8}^{2}$
coefficient also vanishes.

The effective Hamiltonian (\ref{C13}) commutes with
the left
transformation
generators $\hat{L}_{(A)}$ and the right
transformation (spin) generators $\hat{R%
}_{(0,1,M)}$ :
\begin{equation}
\left[ \hat{L}_{(A)},H\right] =\left[ \hat{R}_{(0,1,M)},H\right]
=0\, , \label{F1}
\end{equation}
which ensures that the states (\ref{C16}) are the eigenstates of
the effective Hamiltonian.

The symmetry breaking term does, however, not commute with the
left generators:
\begin{equation}
\left[ \hat{L}_{(Z,\frac{1}{2},M)},M_{SB}\right] \neq 0\, ,
\label{F2}
\end{equation}
and therefore this term mix the states in different
representations $(\Lambda ,M)$.

A new result of this investigation
is the tensor term (\ref{C16}) in Hamiltonian
density operator
(\ref{C15}).
Because of the tensor operator the states
with spin $S>%
\frac{1}{2}$ have quadrupole moments.

Consider finally the energy functional of the quantum skyrmion
in the states of $(\Lambda ,M)$ irrep. The problem is
simplified if the symmetry breaking term that leads to
representation mixing is dropped:
\begin{align}
E(F)& = \frac{C_{2}^{SU(3)}
(\Lambda ,M)-\lambda ^{\prime 2}}{a_{\frac{1}{2}}(F)}%
+\left( \frac{1}{a_{1}(F)}-\frac{1}{a_{\frac{1}{2}}(F)}\right)
S(S+1) \notag \\
& \phantom{W}+\Delta M_{1}+\Delta M_{2}+\Delta M_{3}+M_{cl}\,.
\label{F3}
\end{align}

The variational condition for the energy is:
\begin{equation}
\frac{\delta E(F)}{\delta F}=0\, ,  \label{F4}
\end{equation}
with the usual boundary conditions $F(0)=\pi ,F(\infty )=0$. At
large distances this equation reduces to the asymptotic form
\begin{equation}
\tilde{r}^{2}F^{\prime \prime }+2\tilde{r}F^{\prime }-(2+\tilde{m}^{2}%
\widetilde{r}^{2})F=0\, ,  \label{F5}
\end{equation}
where the quantity $\tilde{m}^{2}$ is defined as:
\begin{align}
\tilde{m}^{2}& = -e^{4}\Bigl(
\frac{1}{4\tilde{a}_{\frac{1}{2}}^{2}(F)}\left(
C_{2}^{SU(3)}(\Lambda ,M)-S(S+1)-\lambda ^{\prime 2}+1\right)
+\frac{2 \, S(S+1)+3}{3 \, \tilde{a}_{1}^{2}(F)} \notag \\
& \phantom{W}+\frac{8\Delta \tilde{M}_{1}+4\Delta \tilde{M}_{3}}{3
\,  \tilde{a}_{1}(F)}+\frac{\Delta \tilde{M}_{3}+2\Delta
\tilde{M}_{2}}{2 \, \tilde{a}_{\frac{1}{2}}(F)}+\frac{1}{\tilde{a}_{1}(F)%
\tilde{a}_{\frac{1}{2}}(F)} \Bigl)\, . \label{F6}
\end{align}
The corresponding asymptotic solution takes the form:
\begin{equation}
F(\tilde{r})=k\left( \frac{\tilde{m}^{2}}{\tilde{r}}+\frac{1}{\tilde{r}^{2}}%
\right) \exp (-\tilde{m}\tilde{r})\, . \label{F7}
\end{equation}
The quantum corrections depends on the irrep $(\lambda,\mu)$ to which
the unitary field $U(\mathbf{x},t)$ belongs as well as on the
state irrep ($\Lambda,M$) and spin $S$.
This bears on the
stability of quantum skyrmion, the requirement of stability of
which is
that the integrals (\ref
{B81}, \ref{B8}) and $\Delta M_{k}$
converge. This requirement is satisfied only if $%
\tilde{m}^{2}>0$. That condition is only
satisfied in the presence of the negative quantum mass corrections $%
\Delta M_{k}$. It is the absence of this term, which leads to the
instability of the skyrmion in the semiclassical approach
\cite{Acus98} in the SU(2) case. Note that in the quantum
treatment \ the chiral angle $F(\tilde{r})$ has the asymptotic
exponential behavior (\ref {F7}) even in the chiral limit.

\begin{acknowledgments}

We would like to thank A. Acus for discussions on group
theoretical problems. Research supported in part by
the Academy of Finland grant number 54038

\end{acknowledgments}

\clearpage

\appendix

\section{}

The functions $C_{\alpha }^{^{\prime }(\bar{A})}(q)$ defined in
(\ref{A2}) are siebenbeins which constitute nonsingular $7\times
7$ matrices. We can introduce the reciprocal functions
$C_{(\bar{B})}^{^{\prime }\alpha }(q)$ by:
\begin{subequations}
\begin{align}
\sum_{\bar{A}}C_{\alpha }^{^{\prime }(\bar{A})}(q)\cdot C_{(\bar{A}%
)}^{^{\prime }\beta }(q)& =\delta _{\alpha \beta }\,,   \\
\sum_{\alpha }C_{\alpha }^{^{\prime }(\bar{A})}(q)\cdot C_{(\bar{B}%
)}^{^{\prime }\alpha }(q)& =\delta _{(\bar{A})(\bar{B})}\,.
\label{P1}
\end{align}
\end{subequations}
Here $(\bar{A})$ and $(\bar{B})$ denote the basis of the
irrep $(1,1)$, with exception for
the state $(0,0,0).$ The $C_{(0)}^{^{\prime }\alpha }(q)$
are not defined.

The properties of the functions
$C_{\alpha }^{\prime (K)}(q)$ follow from $%
\partial _{\alpha }\partial _{\beta }D^{(\lambda ,\mu )}=\partial _{\beta
}\partial _{\alpha }D^{(\lambda ,\mu )}$:\
\begin{equation}
\partial _{\beta }C_{\alpha }^{^{\prime }(K)}(q)-\partial _{\alpha }C_{\beta
}^{^{\prime }(K)}(q)-\sqrt{3}\renewcommand{\arraystretch}{0.9}\left[
\begin{array}{ccc}
\scriptstyle(1,1) & \scriptstyle(1,1) & \scriptstyle(1,1)_{a} \\
\scriptstyle(K^{\prime }) & \scriptstyle(K^{\prime \prime }) & \scriptstyle%
(K)
\end{array}
\right] C_{\beta }^{^{\prime }(K^{\prime })}(q)C_{\alpha }^{^{\prime
}(K^{\prime \prime })}(q)=0\,,
\label{P2}
\end{equation}
and are correct for all states $(K)$ including $(0,0,0)$. The
following properties of the functions $C_{(\bar{K})}^{^{\prime
}\alpha }(q)$ are useful:
\begin{align}
&C_{(\bar{K}^{\prime })}^{^{\prime }\alpha }(q)\partial _{\alpha }C_{(\bar{K%
}^{\prime \prime })}^{^{\prime }\beta }(q)-C_{(\bar{K}^{\prime \prime
})}^{^{\prime }\alpha }(q)\partial _{\alpha }C_{(\bar{K}^{\prime
})}^{^{\prime }\beta }(q)+\sqrt{3}\renewcommand{\arraystretch}{0.9}\left[
\begin{array}{ccc}
\scriptstyle(1,1) & \scriptstyle(1,1) & \scriptstyle(1,1)_{a} \notag \\
\scriptstyle(\bar{K}^{\prime }) & \scriptstyle(\bar{K}^{\prime \prime }) & %
\scriptstyle(\bar{K})
\end{array}
\right] C_{(\bar{K})}^{^{\prime }\beta }(q) \\
&=\sqrt{3}z^{\prime \prime }C_{\alpha }^{^{\prime }(0)}(q)C_{(\bar{K}%
^{\prime })}^{^{\prime }\alpha }(q)C_{(\bar{K}^{\prime \prime })}^{^{\prime
}\beta }(q)-\sqrt{3}z^{\prime }C_{\alpha }^{^{\prime }(0)}(q)C_{(\bar{K}%
^{\prime \prime })}^{^{\prime }\alpha }(q)C_{(\bar{K}^{\prime
})}^{^{\prime }\beta }(q)\,. \label{Cfunc}
\end{align}

In section 4 the right transformation generators (\ref{C6}) are
defined with the following commutation relations:
\begin{align}
\left[ \hat{R}_{(\bar{A}^{\prime })},\hat{R}_{(\bar{A}^{\prime \prime })}%
\right] &=-\sqrt{3}\renewcommand{\arraystretch}{0.9}\left[
\begin{array}{ccc}
\scriptstyle(1,1) & \scriptstyle(1,1) & \scriptstyle(1,1)_{a} \\
\scriptstyle(\bar{A}^{\prime }) & \scriptstyle(\bar{A}^{\prime \prime }) & %
\scriptstyle(\bar{A})
\end{array}
\right] \hat{R}_{(\bar{A})} \notag  \\
& \phantom{W}+\sqrt{3}z^{\prime \prime }\left\{
C_{(\bar{A}^{\prime })}^{\prime \alpha }(q)C_{\alpha }^{\prime
(0)}(q),\hat{R}_{(\bar{A}^{\prime \prime })}\right\}
-\sqrt{3}z^{\prime }\left\{ C_{(\bar{A}^{\prime \prime })}^{\prime
\alpha }(q)C_{\alpha }^{\prime (0)}(q),\hat{R}_{(\bar{A}^{\prime
})}\right\}\ . \notag\ \\
\label{P4}
\end{align}

The SU(2) subalgebra of the generators $\hat{R}_{(0,1,M)}$
satisfies the standard SU(2) commutation relations. These may be
interpreted as spin operators because its acting on unitary field
(\ref{B1}) can be realized as a spatial rotation of skyrmion only:
\begin{equation}
\left[ \hat{R}_{(0,1,M)},A(q)U_{0}(x) A(q)^\dagger\right]
=A(q)\left[ J_{(0,1,M)}^{(1,1)},U_{0}(x)\right]
A^\dagger (q).
\label{P.5}
\end{equation}

The transformation rule for irrep matrices is:
\begin{align}
\left[ \hat{R}_{(\bar{K})},D_{(A)(A^{\prime })}^{(\lambda ,\mu
)}(q)\right] =& \, D_{(A)(A^{\prime \prime })}^{(\lambda ,\mu
)}(q) \genfrac{\langle}{|}{0pt}{}{(\lambda ,\mu )}{ A^{\prime
\prime}} \hspace{0.1cm}J_{(\bar{K})}^{(1,1)}
\genfrac{|}{\rangle}{0pt}{}{(\lambda ,\mu )}{ A^{\prime}}
\notag \\
&-\frac{\sqrt{3}}{2}%
y^{\prime }C_{(\bar{K})}^{^{\prime }\alpha }(q)C_{\alpha
}^{^{\prime }(0)}(q)D_{(A)(A^{\prime })}^{(\lambda ,\mu )}(q)\,.
\label{P6}
\end{align}

The eight left transformation generators are defined as:
\begin{align}
\hat{L}_{(B)} &=\frac{1}{2}\left\{ \hat{R}_{(A)},D_{(A)(B)}^{(1,1)}(-q)%
\right\}  \notag  \\
&=\frac{i}{2}\left\{ p_{\beta }+\lambda iC_{\beta }^{\prime
(0)}(q),K_{(B)}^{\beta }(q)\right\} +\lambda
D_{(0)(B)}^{(1,1)}(-q),  \label{P7}
\end{align}
where
\begin{equation}
K_{(B)}^{\beta }(q)=C_{(\bar{A})}^{\prime \beta }(q)D_{(\bar{A}%
)(B)}^{(1,1)}(-q)\, ,  \label{P8}
\end{equation}
the properties of which follows from (\ref{Cfunc}):
\begin{equation}
K_{(B^{\prime \prime })}^{\beta ^{\prime \prime }}(q)\partial
_{\beta ^{\prime \prime }}K_{(B^{\prime })}^{\beta ^{\prime
}}(q)-K_{(B^{\prime })}^{\beta ^{\prime \prime }}(q)\partial
_{\beta ^{\prime \prime
}}K_{(B^{\prime \prime })}^{\beta ^{\prime }}(q)=\sqrt{3}\renewcommand{%
\arraystretch}{0.9}\left[
\begin{array}{ccc}
\scriptstyle(1,1) & \scriptstyle(1,1) & \scriptstyle(1,1)_{a} \\
\scriptstyle(B^{\prime \prime }) & \scriptstyle(B^{\prime }) & \scriptstyle%
(B)
\end{array}
\right] K_{(B)}^{\beta ^{\prime }}(q)\,.
\end{equation}

By making use of (\ref{P4}) it may be proven that:
\begin{equation}
\left[ \hat{L}_{(B^{\prime })},\hat{L}_{(B^{\prime \prime })}\right] =\sqrt{3%
}\renewcommand{\arraystretch}{0.9}\left[
\begin{array}{ccc}
\scriptstyle(1,1) & \scriptstyle(1,1) & \scriptstyle(1,1)_{a} \\
\scriptstyle(B^{\prime }) & \scriptstyle(B^{\prime \prime }) & \scriptstyle%
(B)
\end{array}
\right] \hat{L}_{(B)}\, .  \label{P9}
\end{equation}
The three right transformation generators or spin operators
$\hat{R}_{(0,1,M)}$ commute with the left transformation generators:
\begin{equation}
\left[ \hat{R}_{(0,1,M)},\hat{L}_{(B)}\right] =0\,.  \label{D10}
\end{equation}

The left transformation rules for the irrep matrices
are:
\begin{align}
\left[ \hat{L}_{(B)},D_{(A^{\prime })(A)}^{(\lambda ,\mu )}(q)\right]
=& \, \left\langle A^{\prime }\left|
\hspace{0.1cm}J_{(B)}^{(1,1)}\hspace{0.05cm%
}\right| A^{\prime \prime }\right\rangle D_{(A^{\prime \prime
})(A)}^{(\lambda ,\mu )}(q)-\frac{\sqrt{3}}{2}y_{A}\cdot
D_{(0)(B)}^{(1,1)}(-q)D_{(A^{\prime })(A)}^{(\lambda ,\mu )}(q)
\notag
\\
&-\frac{\sqrt{3}}{2}y_{A}\cdot C_{\alpha }^{^{\prime }(0)}(q)C_{(\bar{B}%
^{\prime })}^{^{\prime }\alpha }(q)D_{(\bar{B}^{\prime
})(B)}^{(1,1)}(-q)D_{(A^{\prime })(A)}^{(\lambda ,\mu )}(q).
\label{D12}
\end{align}
It is straightforward to derive the following result:
\begin{align}
(-1)^{B}\hat{L}_{(B)}\hat{L}_{(-B)}&  =
(-1)^{\bar{A}}\hat{R}_{(\bar{A})}\hat{R}_{(-\bar{A})}+\lambda
^{^{\prime } 2}-\frac{3}{4} \notag \\
&\phantom{W} -\frac{3}{16}(-1)^{%
\overset{\phantom{i}=}{A}}C_{\alpha }^{^{\prime }(0)}(q)C_{(\overset{%
\phantom{i}=}{A})}^{^{\prime }\alpha }(q)C_{\beta }^{^{\prime }(0)}(q)C_{(-%
\overset{\phantom{i}=}{A})}^{^{\prime }\beta }(q)\, .
\label{lefttrans}
\end{align}

For the derivation of the Lagrangian density
the following expressions are needed:
\begin{align}
& E^{(\bar{A})(\bar{B})}(F)
\,J_{(\bar{A})}^{(1,1)}J_{(\bar{B})}^{(1,1)} \notag \\
& \phantom{W}=-\frac{1}{a_{\frac{1}{2}}(F)}\hat{C}_{2}^{SU(3)}+\left( \frac{1%
}{a_{\frac{1}{2}}(F)}-\frac{1}{a_{1}(F)}\right) \hat{C}^{SU(2)}+\frac{1}{a_{%
\frac{1}{2}}(F)}\left( J_{(0,0,0)}^{(1,1)}\right) ^{2}\, ,
\end{align}
\begin{align}
& E^{(\bar{A})(\bar{B})}(F)
D_{(\bar{B}^{\prime })(\bar{B})}^{I}(\hat x, F(r))\,J_{(%
\bar{A})}^{(1,1)}J_{(\bar{B}^{\prime })}^{(1,1)} \notag \\
& \phantom{W}=-\frac{\cos F}{a_{\frac{1}{2}}(F)}\hat{C}_{2}^{SU(3)}+\left(
\frac{\cos F}{a_{\frac{1}{2}}(F)}-\frac{\cos 2F}{a_{1}(F)}\right) \hat{C}%
^{SU(2)}+\frac{\cos F}{a_{\frac{1}{2}}(F)}\left( J_{(0,0,0)}^{(1,1)}\right)
^{2} \notag \\
& \phantom{WW}+i\left( \frac{\sin 2F}{a_{1}(F)}+\frac{\sin F}{a_{\frac{1}{2}%
}(F)}\right) \left( J_{(0,1,\cdot )}^{(1,1)}\cdot \hat{x}\right) -2\,\frac{%
\sin ^{2}F}{a_{1}(F)}\left( J_{(0,1,\cdot )}^{(1,1)}\cdot \hat{x}\right)
\left( J_{(0,1,\cdot )}^{(1,1)}\cdot \hat{x}\right)\, .
\end{align}
Here $D_{(\bar{B}^{\prime })(\bar{B})}^{I}(\hat x, F(r))$ is
a Wigner matrix of the SU(2). The summation is over SU(2)
representations $I=\frac{1}{2},1$ and the corresponding bases,
\begin{equation}
E^{(\bar{A})(\bar{B})}(F)\renewcommand{\arraystretch}{0.9}\left[
\begin{array}{ccc}
\scriptstyle(1,1) & \scriptstyle(1,1) & \scriptstyle(1,1)_{a} \\
\scriptstyle(\bar{A}) & \scriptstyle(0,1,u) & \scriptstyle(\bar{C})
\end{array}
\right] \,J_{(\bar{B})}^{(1,1)}J_{(\bar{C})}^{(1,1)}=\frac{1}{\sqrt{3}}%
\left( \frac{1}{2a_{\frac{1}{2}}(F)}+\frac{1}{a_{1}(F)}\right)
J_{(0,1,u)}^{(1,1)}\,,
\end{equation}
\begin{align}
& E^{(\bar{A})(\bar{B})}(F)\renewcommand{\arraystretch}{0.9}\left[
\begin{array}{ccc}
\scriptstyle(1,1) & \scriptstyle(1,1) & \scriptstyle(1,1)_{a} \notag \\
\scriptstyle(\bar{A}) & \scriptstyle(0,1,u) & \scriptstyle(\bar{C})
\end{array}
\right] \,D_{(\bar{C}^{\prime })(\bar{C})}^{I}(\hat x, F(r))J_{(\bar{B}%
)}^{(1,1)}J_{(\bar{C}^{\prime })}^{(1,1)} \notag \\
& \phantom{W}=\frac{1}{\sqrt{3}} \Bigl\{-\left[ J_{(0,1,\cdot
)}^{(1,1)}\times \hat{x}\right] _{u}\left( \frac{\sin F}{2a_{\frac{1}{2}}(F)}%
+i\frac{2\sin ^{2}F}{a_{1}(F)}\left( J_{(0,1,\cdot )}^{(1,1)}\cdot \hat{x}%
\right) \right)  \notag \\
& \phantom{WWWWpl}-i\left( \frac{\sin F}{2a_{\frac{1}{2}}(F)}\hat{C}%
_{2}^{SU(3)}-\left( \frac{\sin F}{2a_{\frac{1}{2}}(F)}-\frac{\sin 2F}{%
a_{1}(F)}\right) \hat{C}^{SU(2)}-\frac{\sin F}{2a_{\frac{1}{2}}(F)}\left(
J_{(0,0,0)}^{(1,1)}\right) ^{2}\right) \,\hat{x}_{u} \notag \\
& \phantom{WWWWpl}+\left( \frac{\cos 2F}{a_{1}(F)}+\frac{\cos F}{2a_{\frac{1%
}{2}}(F)}+i\frac{\sin 2F}{a_{1}(F)}\left( J_{(0,1,\cdot )}^{(1,1)}\cdot \hat{%
x}\right) \right) \,J_{(0,1,u)}^{(1,1)}\Bigl\}\,.
\end{align}

\end{document}